\documentclass[twocolumn,showpacs,preprintnumbers,amsmath,amssymb]{revtex4}

\usepackage{graphicx}
\usepackage{dcolumn}
\usepackage{bm}

\begin{document}

\preprint{APS/123-QED}

\title{Proof of Heisenberg's error-disturbance principle}

\author{Seiji Kosugi}
 \email{kosugi@daijo.shukutoku.ac.jp}
\affiliation{%
Junior College Division, Shukutoku University, 6-36-4 Maenocho, Itabashi-ku, Tokyo 174-8631, Japan
}%

\date{\today}

\begin{abstract}
According to Heisenberg's arguments about his error-disturbance relation for electron position measurement, the measurement error of the position of an electron determines its uncertainty just after the measurement.
It is the resolution $\epsilon(x_t)$, i.e., the measurement error of the post-measurement observable $\hat{x}_t$, not the precision $\epsilon(x_0)$, i.e., the measurement error of the pre-measurement observable $\hat{x}_0$ that determines the uncertainty of the observable $\hat{x}_t$.
Therefore, Heisenberg's relation must be interpreted as being between the resolution $\epsilon(x_t)$ and disturbance $\eta(y_0)$.
The magnitude of the disturbance $\eta(y_0)$ is independent of the definition of the measurement value of the observable $\hat{x}_t$.
Heisenberg's error-disturbance relation is proven to hold true in general, when the measurement value of $\hat{x}_t$ is defined to minimize the resolution $\epsilon(x_t)$.
\end{abstract}

\maketitle

In 1927, on the basis of his famous $\gamma$ -ray microscope thought experiment, Heisenberg argued  that the error-disturbance relation (EDR) 
\begin{eqnarray}
\epsilon\eta \sim h,
\end{eqnarray}
holds true between the error $\epsilon$ of position measurement of an electron and the disturbance $\eta$ of the electron momentum \cite{WHZP}.
He further wrote "If there existed experiments which allowed simultaneously a "sharper" determination of $p$ and $q$ than equation (1) permits, then quantum mechanics would be impossible."

His reasoning was as follows: Suppose that the momentum of a free electron is precisely known, while the position is completely unknown.
Then, a subsequent measurement of the position is performed with a very small error $\epsilon$.
Because Heisenberg considered the error $\epsilon$ to be identical to the uncertainty of the electron position just after the measurement \cite{WHPP}, if the electron momentum was not altered drastically by the position measurement, the Kennard--Robertson uncertainty relation \cite{Kennard, Robertson} $\sigma(x_t)\sigma(p_t) \geq \hbar /2$ would not hold for the post-measurement state, with $\sigma(A)$ being the standard deviation of an observable $\hat{A}$.
Thus, quantum mechanics would become inconsistent.
The EDR (1) states that in order not to cause such a contradiction, every position measurement must alter the momentum such that after the measurement our knowledge of the electron motion will remain restricted by the Kennard--Robertson uncertainty relation.

Although Heisenberg demonstrated \cite{WHPP} in a few thought experiments that the uncertainty relation (1) was valid, it has never been proven that relation (1) holds true in general.
Ozawa claimed in 2002 \cite{MOPL1} that there exists a position measurement where the error $\epsilon$ is zero and $\eta < \infty$.
Therefore, Eq.(1) would be invalid for this measurement.
Recently, many experimental tests of Heisenberg's EDR have been performed for projective neutron-spin measurements \cite{Nat} and single-photon polarization measurements \cite{ScR}-\cite{MR}.
These experimental tests demonstrated that Heisenberg's EDR 
\begin{eqnarray}
  \epsilon(x_0) \eta(y_0) \geq \frac{1}{2}|\langle \phi_0 | [\hat{x}_0,\hat{y}_0] | \phi_0 \rangle |
\end{eqnarray}
between the measurement error $\epsilon(x_0)$ of a pre-measurement observable $\hat{x}_0$ and the disturbance $\eta(y_0)$ of another observable $\hat{y}_0$ caused by the $\hat{x}_0$ measurement is violated, where $| \phi_0 \rangle$ is the state vector of a measured object just before the measurement.
A universally valid uncertainty relation derived by Ozawa \cite{MOPL2} 
\begin{eqnarray}
  \epsilon(x_0) \eta(y_0)+\epsilon(x_0)\sigma(y_0) + \sigma(x_0) \eta(y_0) \nonumber  \\ 
\geq \frac{1}{2}|\langle \phi_0 | [\hat{x}_0,\hat{y}_0] | \phi_0 \rangle |
\end{eqnarray}
holds true in general, where $\sigma(x_0)$ and $\sigma(y_0)$ are the standard deviations of observables $\hat{x}_0$ and $\hat{y}_0$, respectively, for the initial state $| \phi_0 \rangle$.

Do these facts demonstrate that Heisenberg's abovementioned argument is wrong? The answer is that Heisenberg's original EDR is not relation (2), although Ozawa and the authors of references \cite{ScR}-\cite{MR} regarded it as such.
The author agrees that inequality (2) is not always true; however, in Heisenberg's original reasoning, the error of position measurement is considered to determine the uncertainty of the electron position after the measurement.
This is also supported by the fact that Heisenberg concluded the uncertainty relation to be valid for electron motion after the measurement in the $\gamma$-ray microscope thought experiment \cite{WHPP}.
W. M. de Muynck also remarked that the inequality does not refer to the past but the future, i.e., to the state of the electron after the measurement \cite{Muyn}.
This fact often remains unnoticed.
Therefore, the measurement error in Heisenberg's EDR must be the one that determines the uncertainty in the electron position immediately after the measurement.
As will be demonstrated later, it is the resolution $\epsilon(x_t)$, not the precision $\epsilon(x_0)$ that determines the uncertainty of the observable $\hat{x}_t$ after the measurement (see Eq.(14)).
Therefore, Heisenberg's original EDR must be interpreted as a relation between the resolution $\epsilon(x_t)$ and the disturbance $\eta(y_0)$:
\begin{eqnarray}
  \epsilon(x_t) \eta(y_0) \geq \frac{1}{2}|\langle \phi_0,\xi_0 | [\hat{x}_t,\hat{y}_t] | \phi_0,\xi_0 \rangle |,
\end{eqnarray}
where $| \xi_0 \rangle$ is the state vector of a probe just before the measurement, and we represent the tensor product $|\phi_0\rangle \otimes |\xi_0 \rangle$ as $|\phi_0, \xi_0 \rangle$.
This relation always holds true, as will be shown later.

Here, we will give a brief description of our measurement model.
We consider the measurement of an observable $\hat{x}_0$ of a microscopic object.
The object interacts with a probe, which is a part of an apparatus, in a time interval $(0,t)$.
Let $\hat{U}$ be a unitary operator representing the time evolution of the object-probe system for this time interval.
Then, the object and probe observables $\hat{x}_t$ and $\hat{X}_t$ after the interaction are given by $\hat{U}^{\dagger} (\hat{x}_0 \otimes \hat{I}) \hat{U}$ and $\hat{U}^{\dagger} (\hat{I} \otimes \hat{X}_0) \hat{U}$, respectively. 
We hereafter abbreviate the tensor product $\hat{x}_0 \otimes \hat{I}$ as $\hat{x}_0$.
The probe is supposed to be prepared in a fixed state $|\xi_0 \rangle$.
After the interaction, the probe observable $\hat{X}_t$ is measured using another measurement apparatus.
Then, using the measurement value $X$ obtained, the measurement values of $\hat{x}_0$ and $\hat{x}_t$ are determined.
It is assumed that the probe observable $\hat{X}_t$ can be precisely measured and the apparatus for measuring $\hat{X}_t$ does not interact with the object system.
Such a measurement is called an indirect measurement model.
Every measurement is statistically equivalent to one using indirect measurement models \cite{MOAP}.

There has been no discussion about the problem of how to determine the measurement value in the indirect measurement model.
The measurement value of the observable $\hat{X}_t$ is usually assumed to be that of the observable $\hat{x}_0$.
However, this assumption has no theoretical justification.
As has been pointed out in a previous paper \cite{SK1} in which the author studied position measurements, the measurement value of $\hat{X}_t$ cannot be considered to be the same as that of $\hat{x}_0$.
In the recent experimental tests of Heisenberg's EDR, the measurement value of $\hat{X}_t$  is also regarded as that of the observable $\hat{x}_0$.

The author's latest study has revealed that the measurement values of $\hat{X}_t$ indicate three types of strange behavior in the case where $\hat{x}_0^2=1$ and $\hat{X}_t^2=1$ \cite{SK3}. This case is analogous to that of the recent experimental tests.
The author believes that the measurement values of $\hat{X}_t$ cannot be regarded equivalent to those of $\hat{x}_0$ for these experimental tests because of these strange behaviors.
For example, for a certain state $|\xi_0 \rangle$ of the probe, the distributions of the measurement values of $\hat{X}_t$ are entirely identical to each other for any object state $|\phi_0 \rangle$.
Because one can obtain no information on the measured object from such measurement values, such an experiment cannot be considered to be a measurement.

In this study, the author defines the measurement value of $\hat{x}_t$ to minimize the measurement error $\epsilon(x_t)$, since the magnitude of the disturbance $\eta(y_0)$ is independent of the definition of the measurement value (see Eq.(17)).
Because the measurement value of the observable $\hat{x}_t$ is calculated using that of the observable $\hat{X}_t$ in the indirect measurement model, a measurement-value operator $(\hat{x}_t)_{\rm{m}}$ whose eigenvalues give the measurement values of $\hat{x}_t$ must be a function of the operator $\hat{X}_t$:
\begin{eqnarray}
  (\hat{x}_t)_{\rm{m}}=f(\hat{X}_t).
\end{eqnarray}

Then, the measurement error of the observable $\hat{x}_t$ is defined as 
\begin{eqnarray}
  \epsilon(x_t) = \langle \phi_0,\xi_0 | \{ (\hat{x}_t)_{\rm m}- \hat{x}_t \}^2 |\phi_0,\xi_0 \rangle ^{1/2}.
\end{eqnarray}
We consider here the case where eigenvalues of the observables $\hat{x}_0$ and $\hat{X}_0$ are continuous:
\begin{eqnarray}
  \hat{x}_0 |x \rangle=x |x \rangle, \; \hat{X}_0 |X \rangle=X |X \rangle.  \nonumber
\end{eqnarray}
For the sake of simplicity, we assume that the eigenvalues $x$ and $X$ are non-degenerate.
Then, using the expressions $\hat{x}_t=\hat{U}^{\dagger}\hat{x}_0\hat{U}$ and $\hat{X}_t=\hat{U}^{\dagger}\hat{X}_0\hat{U}$, we obtain
\begin{eqnarray}
  \epsilon^2(x_t) = \int \{ f(X)-x \}^2 |\langle x,X|\hat{U}|\phi_0,\xi_0 \rangle |^2 dx dX,
\end{eqnarray}
where $\langle x,X|\hat{U}|\phi_0,\xi_0 \rangle$ is a wave function of the object-probe system just after the measurement.

One can express the square of an object wave function $\langle x|\phi_t \rangle_X$ just after the measurement as 
\begin{eqnarray}
  |\langle x|\phi_t \rangle_X|^2 = |\langle x,X|\hat{U}|\phi_0,\xi_0 \rangle |^2 /P(X),
\end{eqnarray}
when the readout value of the probe operator $\hat{X}_t$ is $X$, where $P(X)$ is the probability density for obtaining the readout value $X$:
\begin{eqnarray}
  P(X)=\int |\langle x,X|\hat{U}|\phi_0,\xi_0 \rangle |^2 dx.
\end{eqnarray}
Thus, when the square of the measurement error $\epsilon_X(x_t)$ for the readout value $X$ is defined as
\begin{eqnarray}
  \epsilon^2_{X}(x_t) = \int \{ f(X)-x \}^2 |\langle x|\phi_t \rangle_X |^2 dx, 
\end{eqnarray}
we have 
\begin{eqnarray}
  \epsilon^2(x_t) = \int \epsilon^2_{X}(x_t) P(X)dX.
\end{eqnarray}
When the term $x-f(X)$ is modified into $\{x-\langle x \rangle_X\}-\{\langle x \rangle_X-f(X)\}$ with $\langle x \rangle_X=\int x|\langle x|\phi_t \rangle_X |^2dx$, the following equations are obtained:
\begin{eqnarray}
  \epsilon^2_X(x_t) &=& \sigma^2_{X}(x)+\{ \langle x \rangle_X-f(X) \}^2,    \\
  \sigma^2_X(x) &=& \int \{x-\langle x \rangle_X \}^2|\langle x|\phi_t \rangle_X |^2dx,  \nonumber
\end{eqnarray}
where the quantity $\sigma_X(x)$ is the standard deviation of the object observable $\hat{x}_t$ when the readout is $X$.

Because $\sigma_X(x)$ is independent of the choice of the measurement value $f(X)$, it is clear from Eq.(12) that when the measurement value is defined as
\begin{eqnarray}
  f(X) = \langle x \rangle_X,
\end{eqnarray}
the error $\epsilon_X(x_t)$ has its minimal value:
\begin{eqnarray}
  \epsilon_X(x_t)=\sigma_X(x).
\end{eqnarray}
The same result is obtained using the condition that the functional $\epsilon_X(x_t)$ has a maximum or minimum value for any variation $\delta f(X)$ of the function $f(X)$:
\begin{eqnarray}
  \delta \epsilon^2_X(x_t) =2 \int \{ f(X)-x \} |\langle x|\phi_t \rangle_X |^2 dx \delta f(X)=0. \nonumber
\end{eqnarray}

We have already derived Eq.(12) in a previous paper \cite{SK2}, where we discussed the standard quantum limit for monitoring free-mass position.
In that paper, we concluded that $\epsilon_X(x_t) \ge \sigma_X(x)$.
However, because the measurement value $f(X)$ should be defined as Eq.(13), the measurement error $\epsilon_X(x_t)$ coincides with the standard deviation $\sigma_X(x)$.
Equation (14) indicates clearly that it is the resolution $\epsilon(x_t)$, not the precision $\epsilon(x_0)$ that determines the uncertainty of the observable $\hat{x}_t$ just after the measurement.

It is easy to derive the following equation from definition (13) of the measurement value:
\begin{eqnarray}
  \langle \phi_0,\xi_0|(\hat{x}_t)_{\rm{m}}|\phi_0,\xi_0 \rangle = \langle \phi_0,\xi_0|\hat{x}_t |\phi_0,\xi_0 \rangle , 
\end{eqnarray}
which indicates that unbiased measurements of $\hat{x}_t$ minimize the measurement error $\epsilon(x_t)$.
Because Eq.(15) must be satisfied for any initial object state $|\phi_0 \rangle$, we have 
\begin{eqnarray}
  \langle \xi_0|\{ (\hat{x}_t)_{\rm{m}} -\hat{x}_t \}|\xi_0 \rangle _\kappa = 0,
\end{eqnarray}
where $\langle \cdots | \cdots \rangle _\kappa$ represents the partial inner product over the state space of the probe.

Here, we will derive EDR (4).
The disturbance $\eta(y_0)$ of an object observable $\hat{y}_0$ caused by the $\hat{x}_0$ measurement is defined as
\begin{eqnarray}
 \eta(y_0)=\langle \phi_0,\xi_0 | \{ \hat{y}_t-\hat{y}_0 \}^2 |\phi_0,\xi_0 \rangle ^{1/2}. 
\end{eqnarray}
As in the indirect measurement model, the measurement-value operator $(\hat{x}_t)_{\mathrm{m}}$ is a function of $\hat{X}_t$, we have $[\; (\hat{x}_t)_{\rm m} , \; \hat{y}_t \; ] =0$.
Using the relation 
\begin{eqnarray}
[\; (\hat{x}_t)_{\rm m}-\hat{x}_t, \; \hat{y}_t - \hat{y}_0 \; ] = [\; (\hat{x}_t)_{\rm m}  , \hat{y}_t \; ] - [\; \hat{x}_t, \hat{y}_t \; ] \nonumber \\
- [\; (\hat{x}_t)_{\rm m}-\hat{x}_t, \; \hat{y}_0 \; ],   \nonumber
\end{eqnarray}
and the condition (16) of unbiased measurement, we obtain EDR (4).
Therefore, Heisenberg's original EDR holds true in general. 

When we substitute the position $\hat{q}_0$ and momentum $\hat{p}_0$ operators for $\hat{x}_0$ and $\hat{y}_0$, respectively, we have $[\; \hat{q}_t, \; \hat{p}_t \; ]=\hat{U}^{\dagger}[\; \hat{q}_0, \; \hat{p}_0 \; ]\hat{U}= {\rm i}\hbar$.
Then, we obtain
\begin{eqnarray}
  \epsilon(q_t) \eta(p_0) \geq \frac{\hbar}{2}.
\end{eqnarray}

Moreover, the uncertainty relation 
\begin{eqnarray}
  \epsilon(x_t) \sigma(y_t) \geq \frac{1}{2}|\langle \phi_0,\xi_0 | [\; \hat{x}_t, \; \hat{y}_t \; ] | \phi_0,\xi_0 \rangle |
\end{eqnarray}
holds, because $[\; (\hat{x}_t)_{\rm m}-\hat{x}_t, \; \hat{y}_t \; ] = [\; (\hat{x}_t)_{\rm m}  , \hat{y}_t \; ] - [\; \hat{x}_t, \hat{y}_t \; ] $.
This relation is always true in the same manner as Ozawa's universally valid uncertainty relation, because it has been derived without the condition (16) of unbiased measurement. 
This relation is the same as Mensky's uncertainty relation induced by measurement \cite{Mensky}.

When the eigenvalues of $\hat{x}_0$ and $\hat{X}_0$ are discrete, i.e., 
\begin{eqnarray}
  \hat{x}_0 |x_i \rangle=x_i |x_i \rangle, \; \hat{X}_0 |X_j \rangle=X_j |X_j \rangle,  \nonumber
\end{eqnarray}
the measurement value can be defined in the same manner as the case in which these eigenvalues are continuous.

When the readout of $\hat{X}_t$ is $X_j$, the measurement value is defined as
\begin{eqnarray}
   f(X_j)=\langle x \rangle_{X_j}= \sum_{i} x_i |\langle x_i|\phi_t \rangle_{X_j} |^2.
\end{eqnarray}
Then, the error $\epsilon_{X_j}(x_t)$ of the $\hat{x}_t$ measurement has a minimal value:
\begin{eqnarray}
  \sigma_{X_j}(x)=\{\sum_{i} (x_i-\langle x \rangle_{X_j} )^2 |\langle x_i|\phi_t \rangle_{X_j} |^2\}^{1/2}.
\end{eqnarray}
In this case, it is easy to demonstrate that condition (16) of unbiased measurement is also satisfied and Heisenberg's EDR (4) always holds.

\end{document}